\documentclass[]{interact}

\usepackage{lipsum}
\usepackage{xcolor}
\usepackage{mathtools}

\newcounter{mycounter}

\usepackage{epstopdf}
\usepackage[caption=false]{subfig}

\usepackage[numbers,sort&compress]{natbib}
\bibpunct[, ]{[}{]}{,}{n}{,}{,}
\renewcommand\bibfont{\fontsize{10}{12}\selectfont}

\theoremstyle{plain}

\theoremstyle{definition}

\theoremstyle{remark}

\begin{document}


\title{Progress on artificial flat band systems: classifying, perturbing, applying 
}

\author{
\name{C. Danieli\textsuperscript{a}\thanks{CONTACT: Carlo Danieli (author) -- email: carlo.danieli@cnr.it} and S. Flach\textsuperscript{b}\thanks{CONTACT: Sergej Flach (author) -- email: sflach@ibs.re.kr}}
\affil{
\textsuperscript{a} Institute for Complex Systems, National Research Council (ISC-CNR), Via dei Taurini 19, Rome, 00185, Italy; Dipartimento di Fisica “E.R. Caianiello”, Universit\`a di Salerno, Via Giovanni Paolo II, 132, I-84084 Fisciano (SA), Italy}
\textsuperscript{b}Center for Theoretical Physics of Complex Systems, Institute for Basic Science, Daejeon 34126, Republic of Korea; Basic Science Program, Korea University of Science and Technology (UST), Daejeon 34113, Republic of Korea; Centre for Theoretical Chemistry and Physics, The New Zealand Institute for Advanced
Study (NZIAS), Massey University Albany, Auckland 0745,
New Zealand.
}

\maketitle

\begin{abstract}
We highlight recent progress in the study of artificial flat band systems with a threefold focus. First, we discuss single-particle flat band physics, which has advanced through the design of various flat band generators. These generators rely on the classification of flat bands in terms of compact localized states -- their fundamental building blocks. A related development is the complete real-space description of flat band projectors.
Next, we review studies on perturbations of flat bands, which provide new insights into the effects of disorder and, more importantly, the intricate interplay between many-body interactions and flat band physics.
Finally, we survey the growing number of experimental realizations of flat bands across diverse physical platforms.
\end{abstract}


\clearpage

\section{Introduction}
Periodic lattices featuring dispersionless energy bands -- known as flat bands (FBs) -- have become a cornerstone of modern condensed matter and wave physics. 
In lattices with finite-range hopping and a finite number of bands, a flat band is supported by a macroscopic number of degenerate compact localized states (CLS). Their existence originates from destructive wave interference, a phenomenon ubiquitous across all physical FB systems, which make CLS highly sensitive to perturbations. 
The appeal of FBs largely stems from the fact that a vanishing bandwidth implies zero kinetic energy, making even weak perturbations strongly nonperturbative. 
The ongoing search for novel and unconventional transport phases -- often emerging when flat bands are perturbed -- has been a major driving force behind the recent surge of research in this field. 

The review by Leykam et al.~\cite{leykam2018artificial} in Advances in Physics: X provided a comprehensive overview of FB research up to 2018. At that time, most studies focused on a few prototypical models, such as the Lieb lattice, the diamond chain, and the kagome lattice. These systems served as testbeds to explore the effects of perturbations (e.g., disorder or interactions) and to realize flat bands experimentally using photonic waveguides and exciton–polariton condensates~\cite{leykam2018artificial,leykam2018perspective}.
However, a complete and systematic algebraic classification of flat bands remained largely unexplored, despite several pioneering attempts to develop classification schemes and generator constructions. As a result, subsequent research on projectors, perturbations, and many-body interactions in flat band lattices was still in its early stages, awaiting further development.

Since 2018, the field has expanded rapidly, as reflected in Refs.~\cite{rhim2021singular,vicencio2021photonic,danieli2024flat}. In particular, photonics has emerged as one of the most prolific platforms for flat band studies, owing to its capability for direct structure fabrication and precise parameter control~\cite{vicencio2021photonic,danieli2024flat}. Notably, this growth occurred even though flat bands were originally introduced in condensed matter physics, for example to study exotic ground states of many-body Heisenberg spin systems~\cite{derzhko2015strongly}.

The present short review aims to update and extend Ref.~\cite{leykam2018artificial} by summarizing the major advances in flat band physics achieved since 2018, with particular emphasis on (i) classification schemes, (ii) the many-body interacting regime, and (iii) experimental realizations across diverse physical platforms beyond photonics. We also discuss emerging directions and future perspectives for this rapidly evolving field.

\section{Classifying, generating, projecting}

Flat band lattice models arise from fine-tuning within tight-binding networks. They form continuous families of parameters in this space rather than existing as isolated points. The essential control parameters are simply the amplitudes of the compact localized states (CLS). This finite set of amplitudes defines the algebraic structure of the CLS set, determines the flat band class, and governs the resulting physical properties~\cite{danieli2024flat}.

\begin{figure}[h]
\centering
\includegraphics[width=\columnwidth]{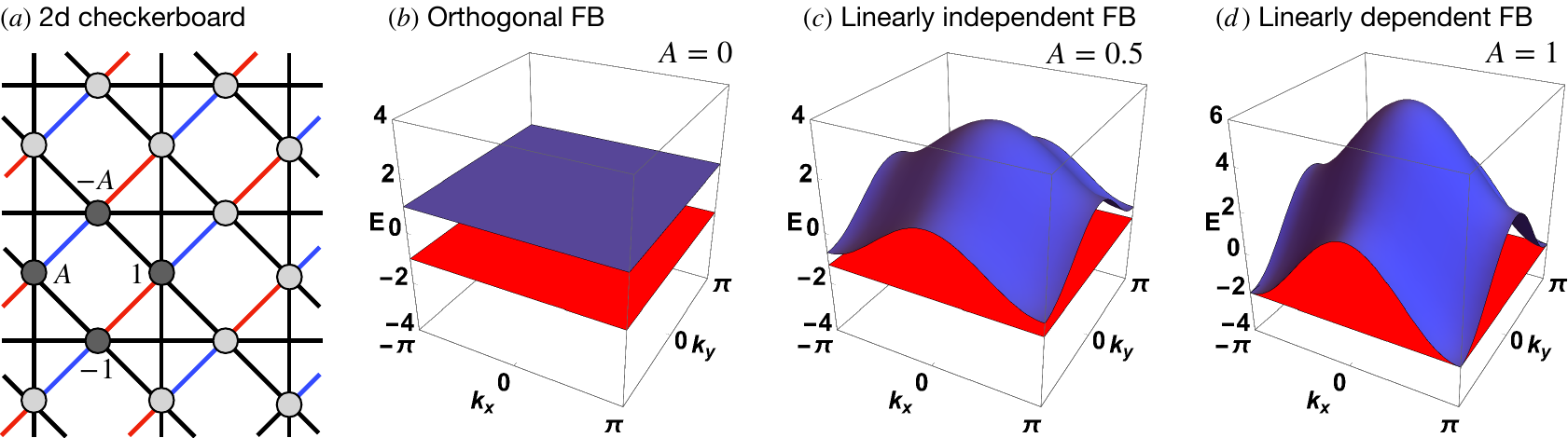}
\caption{(a) Generalized checkerboard lattice. The hopping strengths are equal to $1$ (red lines), $A$ (black lines) and $A^2$ (blue lines). The CLS amplitudes are colored in black. 
(b) Orthogonal flat bands at $E=\pm1$ achieved for $A=0$. 
(c) Linearly independent flat band at $E=-5/4$ (red) gapped from the dispersive band (blue), achieved for $A=0.5$. 
(d) Linearly dependent flat band at $E=-2$ (red) touching the dispersive band (blue), achieved for $A=1$. 
This figure is inspired from some of the results presented in~\cite{kim2025real}.  
} 
\label{fig1}
\end{figure}

The classification of CLS sets is based on two criteria: orthogonality and completeness. A generic choice of CLS amplitudes typically produces a linearly independent but nonorthogonal set. Additional fine-tuning of the CLS amplitudes can render the set orthogonal.
This class includes the notable case of all-bands-flat (ABF) lattices -- such as the Creutz ladder -- where all bands are dispersionless and transport is completely suppressed~\cite{leykam2018artificial}.
Alternatively, a different fine-tuning of the CLS amplitudes can make the set linearly dependent.
This class, known as the singular flat band class~\cite{rhim2021singular}, exists only in dimensions $d \geq 2$ and necessarily features a band touching between the flat band and one or several dispersive bands.
All three classes can coexist within the same CLS controlled parametric family of a flat band lattice and can be accessed by tuning the CLS amplitudes, as illustrated in Fig.~\ref{fig1} for the two-dimensional checkerboard lattice.

The flat band class crucially determines the lattice characteristics in both real and momentum space.
In real space, the flat band projector is a key tool for analyzing lattice perturbations, and its spatial decay depends on the CLS class~\cite{kim2025real}.
Orthogonal flat bands yield strictly compact projectors, whereas nonorthogonal ones produce non-compact projectors due to the overlap between CLSs.
The decay profile also reflects completeness: linearly independent flat bands exhibit exponential decay with algebraic prefactors, while linearly dependent (singular) flat bands display algebraic decay~\cite{kim2025real}.
In momentum space, the energy of an orthogonal flat band is tunable across the spectrum.
Linearly independent flat bands are necessarily gapped from all dispersive bands, whereas linearly dependent flat bands touch at least one dispersive band at one or more points.
These spectral distinctions are illustrated in Fig.~\ref{fig1}(c,d). 

A recent series of CLS-based flat band generators has been introduced and discussed in Ref.~\cite{danieli2024flat}.
The special case of singular flat bands motivated the development of dedicated singular FB generators~\cite{rhim2019classification,hwang2021general}, with particular focus on multifold band-touching points~\cite{graf2021designing}.
A key element common to both the study of flat band projectors~\cite{kim2025real} and singular FB generators~\cite{hwang2021general,rhim2019classification,graf2021designing} is the momentum-space analog of a CLS -- the unnormalized Bloch compact localized state (BCLS) -- and its crucial algebraic structure. 
As a result, elegant and simple parametric FB families can be derived and used for further studies. 
One fresh example is adapted from Ref.~\cite{kim2025real} and shown in Fig.~\ref{fig1}. A $2d$ generalized checkerboard lattice is considered, the CLS on four sites with two having fixed amplitude $1$ and two having tunable amplitude $0 \leq A \leq 1$ is defined, the BCLS is computed, and the generator is applied to arrive at a tunable FB family, linearly independent for all $A$ except for $A=0$ (orthogonal) and $A=1$ (singular). 
Thus, despite substantial progress in the understanding of flat band generation~\cite{leykam2018artificial,danieli2024flat}, the field continues to evolve, producing novel tools and compelling new directions -- for instance, a recently proposed Metropolis-based generator for $d=2$ flat band lattices~\cite{li2025general}.

Even for flat bands, the  Bloch eigenstates  change as one moves through the Brillouin zone. These changes quantify the Quantum Geometric Tensor \cite{yu2025quantumG}. Its real part - the Quantum Metric - defines the {\it distance} between quantum FB states. One of the most profound consequences is that the superfluid stiffness — which determines the transition temperature of a superconductor — is not zero in a flat band, but bounded by the quantum metric.

\section{Many body interactions}

Systematic parametrizations of single-particle FB families are powerful tools for studying the impact of perturbations and for fine-tuning perturbed lattices to exhibit desired responses. Most importantly, the FB projector properties~\cite{kim2025real} provide both quantitative and qualitative insights into the action of external perturbations.
Typical examples include random or correlated disorder and classical nonlinear interactions, which lead to a range of phenomena such as inverse Anderson (metal–insulator) transitions, Bloch oscillations induced by electric fields, and compact breathers and caging effects arising from Kerr nonlinearities — topics covered in detail for photonic applications in Ref.~\cite{danieli2024flat}.

Single-particle CLSs historically served as building blocks for exotic many-body states in interacting quantum systems, such as Heisenberg spin lattices~\cite{derzhko2015strongly}. Their construction allows for new insights into the thermodynamic properties of quantum Heisenberg antiferromagnets in frustrated geometries, for instance the kagome bilayer~\cite{yaremchuk2025frustrated}. Recent developments include both theoretical and experimental studies of entire new classes of quantum materials, collectively referred to as kagome metals~\cite{negi2025magnetic}.

\begin{figure}[h]
\centering
\includegraphics[width=\columnwidth]{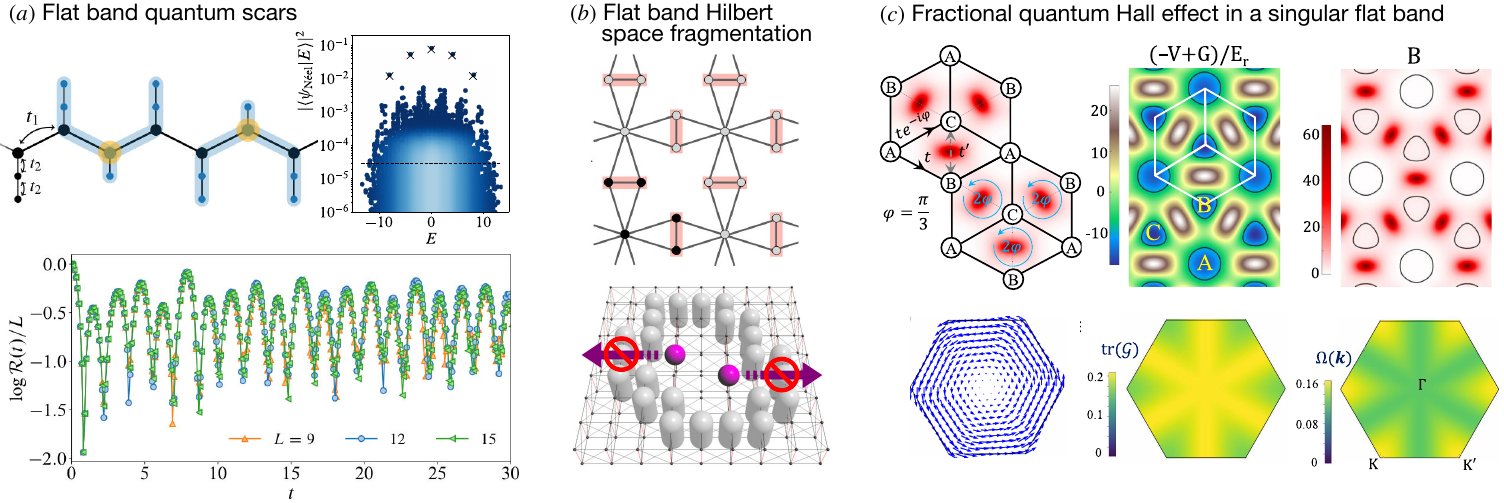}
\caption{
(a)  Eight bands comb structure with flat bands $E=\pm t_2$.  
Projection of the eigenstates onto the N\'eel state, with the scars highlighted with crosses at $E=0,\pm2t_2,\pm4t_2$.  
Return probability $\mathcal{R}$ of a N\'eel state excitation. Reprinted with permission from Ref.~\cite{hart2020compact}. 
(b) Two examples of many-body flat band lattices supporting Hilbert space fragmentation for interacting bosons (top) and hardcore bosons (bottom).  
Reprinted with permission from Refs.~\cite{danieli2020many,lee2024trapping}. 
(c) Top: Dice lattice with pseudomagnetic field (red color); potential landscape and pseudomagnetic field for a possible implementation. 
Bottom: Berry connection; Berry curvature; trace of the quantum metric tensor of the singular flat band.  
Reprinted with permission from Ref.~\cite{yang2025fractional}. 
} 
\label{fig2}
\end{figure}

Non-equilibrium quantum phenomena likewise emerge from many-body interactions in FB lattices. Two prominent examples are {\it quantum many-body scars} and {\it Hilbert-space fragmentation}.
Quantum scars are non-thermalizing many-body states characterized by low entanglement, which have been observed in several interacting FB lattices~\cite{hart2020compact, danieli2021quantum, kuno2021multiple, pelegri2024few}. For example, the system of interacting fermions illustrated in Fig.~\ref{fig2}(a) exhibits scars at five distinct energies. When initialized in an antiferromagnetic N\'eel state (in this case, one particle on every third orange site along the horizontal subchain of the lattice), the return probability shows coherent oscillations rather than rapidly decay to zero~\cite{hart2020compact}.

If FBs emerge from fine-tuning the single-particle sector, one may ask: why not fine-tune the interaction sector as well to obtain new, exotic quantum phases?
Such many-body fine-tuning can yield an extensive set of local commuting operators that fragment the Hilbert space into disconnected subsectors~\cite{kuno2020flat, danieli2020many, lee2024trapping, nicolau2023flat}. This mechanism can completely suppress charge transport along the lattice — a phenomenon known as {\it many-body flat band localization}~\cite{kuno2020flat, danieli2020many, lee2024trapping}.
Two representative systems exhibiting Hilbert-space fragmentation are shown in Fig.~\ref{fig2}(b). In the upper case, the fragmentation arises from finetuned ABF single particle lattices with carefully designed interaction between neighboring sites indicated by colored blocks~\cite{kuno2020flat, danieli2020many}; in the lower case, the shattering of the Hilbert space results from the hard-core boson limit (infinite interaction strength) on a $2d$ cross stitch single particle model with one flat and one dispersive band. 
The figure shows a loop excitation of many CLSs which act as a impenetrable fence for additional hard core bosons trapped inside~\cite{lee2024trapping}.

Further research spans across traditional domains ranging from superconductivity in FB lattices~\cite{bouzerar2025robustness, teeriaho2025coexistence} to paired transport of interacting bosons in Aharonov–Bohm cages~\cite{kolovsky_2023} and the anomalous fractional quantum Hall effect in a singular flat band~\cite{yang2025fractional}. The latter example is illustrated in Fig.~\ref{fig2}(c), showing the Dice lattice dressed by a magnetic field. This model, realizable via a carefully engineered potential landscape, hosts a singular flat band with nontrivial Berry connection, curvature, and quantum metric tensor, which together give rise to the fractional quantum Hall effect.

\section{Experimental applications}

Arrays of optical waveguides have long been considered the optimal platform for experimentally investigating FB phenomena. 
Indeed, the precision offered by femtosecond laser writing allows for the precise engineering of an ever-growing number of more and more complex FB lattices, also including various types of perturbations, {\it e.g.} external fields, periodic driving, interactions~\cite{leykam2018perspective,vicencio2021photonic,danieli2024flat}. 
Nevertheless, significant technological progress has recently enabled other experimental platforms to emerge and stand at the forefront of FB research.  

A fresh example of this is the realization of CLS in electric FB lattices formed by inductors and capacitors. 
In one dimension, these electric circuits can be arranged to form orthogonal FBs (the diamond lattice) and linearly independent FBs (the stub lattice), whose CLS can be continued as exact solutions in the nonlinear regime~\cite{chase2024compact}. Notably, this approach is highly versatile: the circuits in the diamond chain can be reconfigured to emulate the $\pi$-flux within each plaquette, hence yielding the Aharonov-Bohm caging effect~\cite{lape2025realization}.  Moreover, the approach is scalable for generating $2d$ lattices that, for example, (i) feature boundary flat bands with topological spin textures and quantized $\pi$-Berry phases~\cite{biao2024experimental} and (ii) emulate interaction-induced flat-band localization phenomena of two correlated bosons in $1d$ Aharonov-Bohm cages~\cite{zhou2023observation}.

\begin{figure}[h]
\centering
\includegraphics[width=\columnwidth]{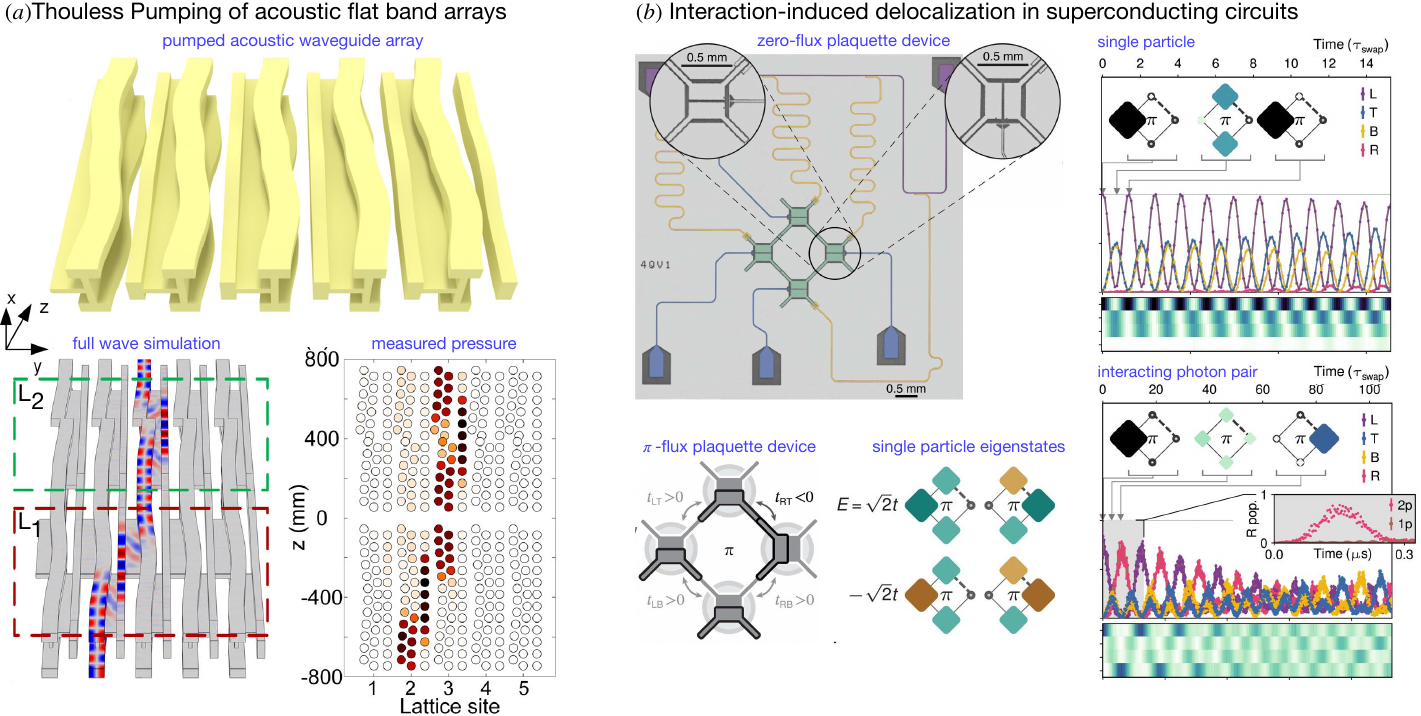}
\caption{
(a) Acoustic platform.   Top: configurations of the acoustic waveguide arrays driven via a pumping loop. 
Bottom: simulated sound wave propagation (left) and measured pressure at discrete locations along the waveguides. 
Reprinted with permission from Ref.~\cite{you2022observation}. 
(b) Superconducting transmon qubit platform. 
Left top: zero-flux plaquette, with zoomed qubit orientations that yield zero (left) and $\pi$-flux (right) plaquette. 
Left bottom: detailed hopping profile of the $\pi$-flux plaquette (left) and single particle eigenbasis (right). 
Right: average qubit populations in the different qubits of the plaquette as function of time for single photon particle (top) and two interacting photon particles (bottom). 
Reprinted with permission from Ref.~\cite{martinez2023interaction}.  
} \label{fig3}
\end{figure}

Acoustic systems also offer a compelling platform for FBs. 
Examples include the experimental realization of CLSs in singular FB lattices inscribed in acoustic plates~\cite{riva2025creating}, and of hexagonic FB lattices created in phononic metamaterials by carefully assembling disks with magnetic couplings~\cite{samak2024direct}. Notably, in the latter example, the magnetic couplings can also be additionally fine-tuned to render all bands flat. 
Furthermore, acoustic systems offer ways to realize topological phenomena -- {\it e.g.} the non-Abelian Thouless pumping in FB systems which was originally proposed in Ref.~\cite{brosco2021nonabelian} for photonic lattices.  
As shown in Fig~\ref{fig3}(a),  non-Abelian pumping can be realized using an array of modulated acoustic waveguides arranged as a generalized one-dimensional Lieb lattice featuring two zero-energy FBs~\cite{you2022observation}. 
The typical displacing and shuffling the CLSs throughout the network, a hallmark of this phenomenon, is shown both numerically (bottom left) and experimentally (bottom right).

As remarked by these experimental results as well as by those in the previous sections, ABF lattices attract significant interest. Indeed, their fully quenched kinetic energy enables exotic phenomena that go beyond traditional FBs. For instance, these systems exhibit the complete absence of single particle transport. 
However, this effect reversed due to Hubbard interactions, which induce two-particle paired transport~\cite{leykam2018artificial,danieli2021quantum,kolovsky_2023}. 
Such a breakdown of full halt of transport has been recently verified using Rydberg atoms in optical arrays~\cite{chen2025interaction},  with a synthetic lattice of  ultracold $\prescript{87}{}{\text{Rb}}$ atoms~\cite{li2025engineering}, and with superconducting transmon qubits~\cite{martinez2023interaction}. 
Fig~\ref{fig3}(b) showcases the latter case by displaying on the left the $\pi$-flux rhombic plaquette of qubits, which is obtained by properly orienting one of the four qubits. 
The top-right panel demonstrates the single particle caging, where a particle starting from the left qubit over time never occupies the right one. 
The bottom-right panel instead shows the caging breakdown for two interacting photons.

These experiments indicate the rapid development of the field of flat bands towards practical technologies, beyond foundational studies.   
Flat bands have been recently used for the realization of lasing using silicon waveguide-integrated metasurfaces~\cite{eyvazi2025flat} and ultra-compact cavities~\cite{cui2025ultracompact}.
Another exciting direction is the observation of topological flat bands nanosheets using photoelectric detections, a result which open venues for new material platform for organic photodetectors~\cite{shi2025characterizing}. 
The field has also seen an avalanche of novel results on kagome metals~\cite{geng2025experimental,oh2025hund,luo2025robust}, which include unusual magnetism~\cite{thomas2025unusual}, high-order topological phases~\cite{lin2025spin} and coherent charge transport phenomena~\cite{guo2025many}. These findings remark this as an intriguing area for the design of unusual quantum states and novel phenomena, promising far-reaching applications in material science. We refer the readers to the following review~\cite{disante2026kagomeM} for more insights.

\section{Conclusions and outlook}

Over the past decade, the field of flat band (FB) physics has matured from a collection of isolated model studies into a coherent and rapidly expanding research frontier that bridges condensed matter, photonics, acoustics, electric circuits, and quantum technologies. The recognition that compact localized states (CLS) form the universal backbone of all flat band systems has enabled a unified algebraic framework that links lattice geometry, interference patterns, and spectral properties. This paradigm shift — from analyzing specific lattices toward understanding the algebraic structure and classification of CLS families — has opened a systematic route for constructing, manipulating, and fine-tuning flat bands across dimensions and physical platforms.

The classification of FBs into orthogonal, linearly independent, and singular  families, together with the associated projector formalism, now provides a clear correspondence between real- and momentum-space features. This structural understanding has demystified phenomena such as algebraic versus exponential localization, band touching, and tunable degeneracies. Moreover, the development of analytical and numerical flat band generators has transformed FB design from an art into an algorithmic process. These tools not only simplify the exploration of parameter spaces but also serve as essential stepping stones for investigating perturbations, nonlinearities, and disorder. The emergence of machine-inspired approaches, such as Metropolis-type algorithms, signals the beginning of automated flat band discovery across higher-dimensional and topologically enriched systems.

Equally transformative has been the exploration of many-body effects in flat band lattices. By suppressing kinetic energy, FBs amplify the role of interactions, giving rise to rich nonperturbative phenomena: quantum scars, Hilbert-space fragmentation, many-body localization without disorder, and interaction-induced transport in otherwise caged systems. The conceptual symmetry between single-particle fine-tuning and many-body fine-tuning illustrates how FBs serve as fertile ground for engineering exotic quantum phases. These advances connect to fundamental questions about ergodicity, thermalization, and the emergence of nontrivial correlations in constrained quantum systems—linking FB physics to the broader landscape of nonequilibrium quantum dynamics.

On the experimental front, remarkable progress has expanded the scope of accessible FB platforms. Photonic waveguide arrays remain a benchmark for precision control and visualization of CLS dynamics, yet complementary platforms—acoustic metamaterials, electric circuits, superconducting qubits, and ultracold atoms—have demonstrated equivalent or even superior tunability. Experiments now routinely probe interaction-driven delocalization, topological pumping, and fractional quantum Hall effects within flat or singular bands. The convergence of these diverse implementations establishes FBs as a universal motif rather than a platform-specific curiosity. Importantly, the cross-fertilization between theory and experiment has accelerated the pace of discovery, revealing FB signatures even in correlated electron materials such as kagome metals.

Looking ahead, flat band research is poised to enter a new integrative phase. Future directions include the design of higher-order topological FBs, controlled inclusion of spin–orbit coupling, and the exploration of nonequilibrium and driven-dissipative regimes where flatness competes with gain and loss. The marriage of algebraic design principles with experimental scalability promises not only deeper theoretical understanding but also tangible applications—from robust photonic circuitry to designer quantum matter with controllable interactions. As a unifying framework for localization, topology, and correlation, flat band physics continues to exemplify how elegant mathematical structure can guide and inspire the search for novel physical realities.

\section*{Acknowledgements}
We thank Alexei Andreanov and Yeongjun Kim for helpful discussions.
C.D. thanks the Center for Theoretical Physics of Complex Systems at the Institute for Basic Sciences, Daejeon, Korea for its generous hospitality during the period when this work was partially completed.

\section*{Disclosure statement}
The authors declare no conflict of interest.

\section*{Funding}
S.F. was supported by the Institute for Basic
Science through Project Code (No. IBS-R024-D1). 
C.D. was supported by the PNRR MUR project CN 00000013-ICSC and the PNRR MUR project PE0000023-NQSTI - TOPQIN.

\bibliographystyle{tfq}
\bibliography{FB2025}

@article{leykam2018artificial,
 author = {Leykam, Daniel and Andreanov, Alexei and Flach, Sergej},
 doi = {10.1080/23746149.2018.1473052},
 journal = {Adv. Phys.: X},
 number = {1},
 pages = {1473052},
 publisher = {Taylor \& Francis},
 title = {Artificial flat band systems: from lattice models to experiments},
 url = {https://doi.org/10.1080/23746149.2018.1473052},
 volume = {3},
 year = {2018},
}

@article{leykam2018perspective,
 author = {Leykam, Daniel and Flach, Sergej},
 doi = {10.1063/1.5034365},
 journal = {APL Photonics},
 number = {7},
 pages = {070901},
 title = {Perspective: Photonic flatbands},
 url = {https://doi.org/10.1063/1.5034365},
 volume = {3},
 year = {2018},
}

@article{danieli2024flat,
url = {https://doi.org/10.1515/nanoph-2024-0135},
title = {Flat band fine-tuning and its photonic applications},
author = {Carlo Danieli and Alexei Andreanov and Daniel Leykam and Sergej Flach},
pages = {3925--3944},
volume = {13},
number = {21},
journal = {Nanophotonics},
doi = {doi:10.1515/nanoph-2024-0135},
year = {2024}
}

@article{derzhko2015strongly,
 author = {Derzhko, Oleg and Richter, Johannes and Maksymenko, Mykola},
 doi = {10.1142/S0217979215300078},
 journal = {Int. J. Mod. Phys. B},
 number = {12},
 pages = {1530007},
 title = {Strongly correlated flat-band systems: The route from Heisenberg spins to Hubbard electrons},
 url = {http://www.worldscientific.com/doi/abs/10.1142/S0217979215300078},
 volume = {29},
 year = {2015},
}

@article{vicencio2021photonic,
  author = {Rodrigo  Poblete},
  title = {Photonic flat band dynamics},
  journal = {Advances in Physics: X},
  volume = {6},
  number = {1},
  pages = {1878057},
  year = {2021},
  publisher = {Taylor & Francis},
  doi = {10.1080/23746149.2021.1878057},
  url = {https://doi.org/10.1080/23746149.2021.1878057}
}

@article{kim2025real,
  title={Real space decay of flat band projectors from compact localized states},
  author={Kim, Yeongjun and Flach, Sergej and Andreanov, Alexei},
  journal={arXiv:2510.17258},
  year={2025}, 
  url={https://arxiv.org/abs/2510.17258}
}

@article{hwang2021general,
  title = {General construction of flat bands with and without band crossings based on wave function singularity},
  author = {Hwang, Yoonseok and Rhim, Jun-Won and Yang, Bohm-Jung},
  journal = {Phys. Rev. B},
  volume = {104},
  issue = {8},
  pages = {085144},
  numpages = {16},
  year = {2021},
  month = {Aug},
  publisher = {American Physical Society},
  doi = {10.1103/PhysRevB.104.085144},
  url = {https://link.aps.org/doi/10.1103/PhysRevB.104.085144}
}

@article{graf2021designing,
  title = {Designing flat-band tight-binding models with tunable multifold band touching points},
  author = {Graf, Ansgar and Pi\'echon, Fr\'ed\'eric},
  journal = {Phys. Rev. B},
  volume = {104},
  issue = {19},
  pages = {195128},
  numpages = {21},
  year = {2021},
  month = {Nov},
  publisher = {American Physical Society},
  doi = {10.1103/PhysRevB.104.195128},
  url = {https://link.aps.org/doi/10.1103/PhysRevB.104.195128}
}

@article{rhim2019classification,
 author = {Rhim, Jun-Won and Yang, Bohm-Jung},
 title = {Classification of flat bands according to the band-crossing singularity of Bloch wave functions},
 journal = {Phys. Rev. B},
 volume = {99},
 issue = {4},
 pages = {045107},
 numpages = {21},
 year = {2019},
 month = jan,
 publisher = {American Physical Society},
 doi = {10.1103/PhysRevB.99.045107},
 url = {https://link.aps.org/doi/10.1103/PhysRevB.99.045107},
}

@article{rhim2021singular,
 author = {Rhim, Jun-Won and Yang, Bohm-Jung},
 title = {Singular flat bands},
 journal = {Advances in Physics: X},
 volume = {6},
 number = {1},
 pages = {1901606},
 year = {2021},
 publisher = {Taylor \& Francis},
 doi = {10.1080/23746149.2021.1901606},
 url = {https://doi.org/10.1080/23746149.2021.1901606},
}

@article{li2025general,
  title = {General Method to Construct Flat Bands in Two-Dimensional Lattices},
  author = {Li, H. T. and Ji, T. Z. and Yan, R. G. and Fan, W. L. and Zhang, Z. X. and Sun, L. and Miao, B. F. and Chen, G. and Wan, X. G. and Ding, H. F.},
  journal = {Phys. Rev. Lett.},
  volume = {134},
  issue = {7},
  pages = {076402},
  numpages = {7},
  year = {2025},
  month = {Feb},
  publisher = {American Physical Society},
  doi = {10.1103/PhysRevLett.134.076402},
  url = {https://link.aps.org/doi/10.1103/PhysRevLett.134.076402}
}

@article{danieli2020many,
  title = {Many-body flatband localization},
  author = {Danieli, Carlo and Andreanov, Alexei and Flach, Sergej},
  journal = {Phys. Rev. B},
  volume = {102},
  issue = {4},
  pages = {041116},
  numpages = {6},
  year = {2020},
  month = {Jul},
  publisher = {American Physical Society},
  doi = {10.1103/PhysRevB.102.041116},
  url = {https://link.aps.org/doi/10.1103/PhysRevB.102.041116}
}

@article{kuno2020flat,
 author = {Kuno, Yoshihito and Orito, Takahiro and Ichinose, Ikuo},
 doi = {10.1088/1367-2630/ab6352},
 url = {https://doi.org/10.1088/1367-2630/ab6352},
 year = {2020},
 month = jan,
 publisher = {{IOP} Publishing},
 volume = {22},
 number = {1},
 pages = {013032},
 title = {Flat-band many-body localization and ergodicity breaking in the Creutz ladder},
 journal = {New J. Phys.}
}

@article{danieli2021quantum,
 author = {Danieli, Carlo and Andreanov, Alexei and Mithun, Thudiyangal and Flach, Sergej},
 title = {Quantum caging in interacting many-body all-bands-flat lattices},
 journal = {Phys. Rev. B},
 volume = {104},
 issue = {8},
 pages = {085132},
 numpages = {10},
 year = {2021},
 month = aug,
 publisher = {American Physical Society},
 doi = {10.1103/PhysRevB.104.085132},
 url = {https://link.aps.org/doi/10.1103/PhysRevB.104.085132},
}

@article{pelegri2024few,
  title = {Few-body bound topological and flat-band states in a Creutz ladder},
  author = {Pelegr\'{\i}, G. and Flannigan, S. and Daley, A. J.},
  journal = {Phys. Rev. B},
  volume = {109},
  issue = {23},
  pages = {235412},
  numpages = {11},
  year = {2024},
  month = {Jun},
  publisher = {American Physical Society},
  doi = {10.1103/PhysRevB.109.235412},
  url = {https://link.aps.org/doi/10.1103/PhysRevB.109.235412}
}

@article{kolovsky_2023,
  title = {Conductance transition with interacting bosons in an Aharonov-Bohm cage},
  author = {Kolovsky, A. R. and Muraev, P. S. and Flach, S.},
  journal = {Phys. Rev. A},
  volume = {108},
  issue = {1},
  pages = {L010201},
  numpages = {5},
  year = {2023},
  month = {Jul},
  publisher = {American Physical Society},
  doi = {10.1103/PhysRevA.108.L010201},
  url = {https://link.aps.org/doi/10.1103/PhysRevA.108.L010201}
}

@article{lee2024trapping,
  title = {Trapping hard-core bosons in flat-band lattices},
  author = {Lee, Sanghoon and Andreanov, Alexei and Sedrakyan, Tigran and Flach, Sergej},
  journal = {Phys. Rev. B},
  volume = {109},
  issue = {24},
  pages = {245137},
  numpages = {9},
  year = {2024},
  month = {Jun},
  publisher = {American Physical Society},
  doi = {10.1103/PhysRevB.109.245137},
  url = {https://link.aps.org/doi/10.1103/PhysRevB.109.245137}
}

@article{hart2020compact,
 author = {Hart, Oliver and De Tomasi, Giuseppe and Castelnovo, Claudio},
 title = {From compact localized states to many-body scars in the random quantum comb},
 journal = {Phys. Rev. Research},
 volume = {2},
 issue = {4},
 pages = {043267},
 numpages = {10},
 year = {2020},
 month = nov,
 publisher = {American Physical Society},
 doi = {10.1103/PhysRevResearch.2.043267},
 url = {https://link.aps.org/doi/10.1103/PhysRevResearch.2.043267},
}

@article{kuno2021multiple,
  title = {Multiple quantum scar states and emergent slow thermalization in a flat-band system},
  author = {Kuno, Yoshihito and Mizoguchi, Tomonari and Hatsugai, Yasuhiro},
  journal = {Phys. Rev. B},
  volume = {104},
  issue = {8},
  pages = {085130},
  numpages = {12},
  year = {2021},
  month = {Aug},
  publisher = {American Physical Society},
  doi = {10.1103/PhysRevB.104.085130},
  url = {https://link.aps.org/doi/10.1103/PhysRevB.104.085130}
}

@article{nicolau2023flat,
  title = {Flat band induced local Hilbert space fragmentation},
  author = {Nicolau, Eloi and Marques, Anselmo M. and Dias, Ricardo G. and Ahufinger, Ver\`onica},
  journal = {Phys. Rev. B},
  volume = {108},
  issue = {20},
  pages = {205104},
  numpages = {9},
  year = {2023},
  month = {Nov},
  publisher = {American Physical Society},
  doi = {10.1103/PhysRevB.108.205104},
  url = {https://link.aps.org/doi/10.1103/PhysRevB.108.205104}
}

@article{yaremchuk2025frustrated,
  title = {Frustrated kagome-lattice bilayer quantum Heisenberg antiferromagnet},
  author = {Yaremchuk, Dmytro and Hutak, Taras and Baliha, Vasyl and Krokhmalskii, Taras and Derzhko, Oleg and Schnack, J\"urgen and Richter, Johannes},
  journal = {Phys. Rev. B},
  volume = {112},
  issue = {2},
  pages = {024402},
  numpages = {17},
  year = {2025},
  month = {Jul},
  publisher = {American Physical Society},
  doi = {10.1103/1rzb-s69p},
  url = {https://link.aps.org/doi/10.1103/1rzb-s69p}
}

@article{teeriaho2025coexistence,
  title = {Coexistence of ergodic and nonergodic behavior and level spacing statistics in a one-dimensional model of a flat band superconductor},
  author = {Teeriaho, Meri and Linho, Ville-Vertti and Swaminathan, Koushik and Peotta, Sebastiano},
  journal = {Phys. Rev. Res.},
  volume = {7},
  issue = {1},
  pages = {013318},
  numpages = {17},
  year = {2025},
  month = {Mar},
  publisher = {American Physical Society},
  doi = {10.1103/PhysRevResearch.7.013318},
  url = {https://link.aps.org/doi/10.1103/PhysRevResearch.7.013318}
}

@article{yang2025fractional,
  title = {Fractional Quantum Anomalous Hall Effect in a Singular Flat Band},
  author = {Yang, Wenqi and Zhai, Dawei and Tan, Tixuan and Fan, Feng-Ren and Lin, Zuzhang and Yao, Wang},
  journal = {Phys. Rev. Lett.},
  volume = {134},
  issue = {19},
  pages = {196501},
  numpages = {7},
  year = {2025},
  month = {May},
  publisher = {American Physical Society},
  doi = {10.1103/PhysRevLett.134.196501},
  url = {https://link.aps.org/doi/10.1103/PhysRevLett.134.196501}
}

@article{bouzerar2025robustness,
  title = {Robustness of flat band superconductivity against disorder in a two-dimensional Lieb lattice model},
  author = {Bouzerar, G. and Thumin, M.},
  journal = {Phys. Rev. B},
  volume = {111},
  issue = {2},
  pages = {L020506},
  numpages = {6},
  year = {2025},
  month = {Jan},
  publisher = {American Physical Society},
  doi = {10.1103/PhysRevB.111.L020506},
  url = {https://link.aps.org/doi/10.1103/PhysRevB.111.L020506}
}

@article{chase2024compact,
  title = {Compact localized states in electric circuit flat-band lattices},
  author = {Chase-Mayoral, Carys and English, L. Q. and Lape, Noah and Kim, Yeongjun and Lee, Sanghoon and Andreanov, Alexei and Flach, Sergej and Kevrekidis, P. G.},
  journal = {Phys. Rev. B},
  volume = {109},
  issue = {7},
  pages = {075430},
  numpages = {9},
  year = {2024},
  month = {Feb},
  publisher = {American Physical Society},
  doi = {10.1103/PhysRevB.109.075430},
  url = {https://link.aps.org/doi/10.1103/PhysRevB.109.075430}
}

@article{lape2025realization,
  title={Realization and characterization of an all-bands-flat electrical lattice},
  author={Lape, Noah and Diubenkov, Simon and English, LQ and Kevrekidis, PG and Andreanov, Alexei and Kim, Yeongjun and Flach, Sergej},
  journal={arXiv:2508.13571},
  year={2025}, 
  url={https://arxiv.org/abs/2508.13571}
}

@article{martinez2023interaction,
  title={Flat-band localization and interaction-induced delocalization of photons},
  author={Martinez, Jeronimo G.C. and Chiu, Christie S. and Smitham, Basil M. and Houck, Andrew A.},
  journal={Science Advances},
  volume={9},
  number={50},
  pages={eadj7195},
  year={2023},
  publisher={American Association for the Advancement of Science},
  doi = {10.1126/sciadv.adj7195},
  url = {https://www.science.org/doi/10.1126/sciadv.adj7195}
}

@article{chen2025interaction,
  title={Interaction-driven breakdown of Aharonov--Bohm caging in flat-band Rydberg lattices},
  author={Chen, Tao and Huang, Chenxi and Velkovsky, Ivan and Ozawa, Tomoki and Price, Hannah and Covey, Jacob P and Gadway, Bryce},
  journal={Nature Physics},
  volume={21},
  number={2},
  pages={221--227},
  year={2025},
  url={https://doi.org/10.1038/s41567-024-02714-7},
  doi={10.1038/s41567-024-02714-7}
      
  }

@article{brosco2021nonabelian,
	title = {Non-Abelian Thouless pumping in a photonic lattice},
	author = {Brosco, Valentina and Pilozzi, Laura and Fazio, Rosario and Conti, Claudio},
	journal = {Phys. Rev. A},
	volume = {103},
	issue = {6},
	pages = {063518},
	numpages = {10},
	year = {2021},
	month = {Jun},
	publisher = {American Physical Society},
	doi = {10.1103/PhysRevA.103.063518},
	url = {https://link.aps.org/doi/10.1103/PhysRevA.103.063518}
}

@article{you2022observation,
  title = {Observation of Non-Abelian Thouless Pump},
  author = {You, Oubo and Liang, Shanjun and Xie, Biye and Gao, Wenlong and Ye, Weimin and Zhu, Jie and Zhang, Shuang},
  journal = {Phys. Rev. Lett.},
  volume = {128},
  issue = {24},
  pages = {244302},
  numpages = {6},
  year = {2022},
  month = {Jun},
  publisher = {American Physical Society},
  doi = {10.1103/PhysRevLett.128.244302},
  url = {https://link.aps.org/doi/10.1103/PhysRevLett.128.244302}
}

@article{cui2025ultracompact,
  title={Ultracompact multibound-state-assisted flat-band lasers},
  author={Cui, Jieyuan and Han, Song and Zhu, Bofeng and Wang, Chongwu and Chua, Yunda and Wang, Qian and Li, Lianhe and Davies, Alexander Giles and Linfield, Edmund Harold and Wang, Qi Jie},
  journal={Nature Photonics},
  volume={19},
  number={6},
  pages={643--649},
  year={2025},
  url={https://doi.org/10.1038/s41566-025-01665-6},
  doi={10.1038/s41566-025-01665-6}
}

@article{eyvazi2025flat,
  title={Flat-Band Lasing in Silicon Waveguide-Integrated Metasurfaces},
  author={Eyvazi, Sioneh and Mamonov, Evgeny A and Heilmann, Rebecca and Cuerda, Javier and Torma, Paivi},
  journal={ACS photonics},
  volume={12},
  number={3},
  pages={1570--1578},
  year={2025},
  doi={doi: 10.1021/acsphotonics.4c02332},
  url={https://doi.org/10.1021/acsphotonics.4c02332}
}

@article{lin2025spin,
    author = {Lin, Qing and Zhou, Pan and Peng, Xiaoning and Sun, Lizhong},
    title = {Spin-polarized flatband and second-order topological phases in fluorinated C3N},
    journal = {Applied Physics Letters},
    volume = {127},
    number = {10},
    pages = {103103},
    year = {2025},
    month = {09},
    issn = {0003-6951},
    doi = {10.1063/5.0279685},
    url = {https://doi.org/10.1063/5.0279685},
    eprint = {https://pubs.aip.org/aip/apl/article-pdf/doi/10.1063/5.0279685/20692123/103103_1_5.0279685.pdf},
}

@article{geng2025experimental,
  title={Experimental realization of dice-lattice flat band at the Fermi level in layered electride YCl},
  author={Geng, Songyuan and Wang, Xin and Guo, Risi and Qiu, Chen and Chen, Fangjie and Wang, Qun and Li, Kangjie and Hao, Peipei and Liang, Hanpu and Huang, Yang and others},
  journal={arXiv:2508.21311},
  year={2025}, 
  url={https://arxiv.org/abs/2508.21311}
}

@article{shi2025characterizing,
author = {Shi, Yanshu and Zhu, Jianhua and Wang, Xuekun and Wang, Yiqian and Song, Qinghao and Chen, Lei and Song, Yumin and Wu, Wei and Guo, Tingting},
title = {Characterizing Topological Flat Bands in Tin-Phthalocyanine Nanosheets Using Photoelectric Detections},
journal = {ACS Photonics},
volume = {12},
number = {7},
pages = {3412-3420},
year = {2025},
doi = {10.1021/acsphotonics.4c02491},
URL = {https://doi.org/10.1021/acsphotonics.4c02491}
}

@article{thomas2025unusual,
  title={Unusual 5f magnetism in new kagome material UV6Sn6},
  author={Thomas, SM and Kengle, CS and Simeth, W and Lim, Chan-young and Riedel, ZW and Allen, K and Schmidt, A and Ruf, M and Gim, Seonggeon and Thompson, JD and others},
  journal={npj Quantum Materials},
  volume={10},
  number={1},
  pages={66},
  year={2025},
  url={https://doi.org/10.1038/s41535-025-00783-2}, 
  doi={10.1038/s41535-025-00783-2}
}

@article{riva2025creating,
  title={Creating compact localized modes for robust sound transport via singular flat band engineering},
  author={Riva, Emanuele and Bellinzoni, Federico and Braghin, Francesco},
  journal={Communications Physics},
  volume={8},
  number={1},
  pages={255},
  year={2025},
  publisher={Nature Publishing Group UK London}, 
  url={https://doi.org/10.1038/s42005-025-02158-2},
  doi={10.1038/s42005-025-02158-2}
}

@article{oh2025hund,
  title={Hund flat band in a frustrated spinel oxide},
  author={Oh, Dongjin and Hampel, Alexander and Wakefield, Joshua P and Moen, Peter and Smit, Steef and Luo, Xiangyu and Zonno, Marta and Gorovikov, Sergey and Leandersson, Mats and Polley, Craig and others},
  journal={arXiv:2502.07234},
  year={2025}, 
  url={https://arxiv.org/abs/2502.07234}
}

@article{samak2024direct,
  title = {Direct Observation of All-Flat Bands Phononic Metamaterials},
  author = {Samak, Mahmoud M. and Bilal, Osama R.},
  journal = {Phys. Rev. Lett.},
  volume = {133},
  issue = {26},
  pages = {266101},
  numpages = {6},
  year = {2024},
  month = {Dec},
  publisher = {American Physical Society},
  doi = {10.1103/PhysRevLett.133.266101},
  url = {https://link.aps.org/doi/10.1103/PhysRevLett.133.266101}
}

@article{biao2024experimental,
  title = {Experimental observation of boundary flat bands with topological spin textures},
  author = {Biao, Yuanchuan and Yan, Zhongbo and Yu, Rui},
  journal = {Phys. Rev. B},
  volume = {110},
  issue = {24},
  pages = {L241110},
  numpages = {7},
  year = {2024},
  month = {Dec},
  publisher = {American Physical Society},
  doi = {10.1103/PhysRevB.110.L241110},
  url = {https://link.aps.org/doi/10.1103/PhysRevB.110.L241110}
}

@article{luo2025robust,
  title={Robust Orbital-Selective Flat Bands in Transition-Metal Oxychlorides},
  author={Luo, Xiangyu and Zullo, Ludovica and Patel, Sahaj and Oh, Dongjin and Song, Qian and Kundu, Asish K and Rajapitamahuni, Anil and Vescovo, Elio and Olszowska, Natalia and Kurleto, Rafal and others},
  journal={arXiv:2510.15080},
  year={2025}, 
url={https://arxiv.org/abs/2510.15080}
}

@article{li2025engineering,
  title={Engineering topological chiral transport in a flat-band lattice of ultracold atoms},
  author={Li, Hang and Liang, Qian and Dong, Zhaoli and Wang, Hongru and Yi, Wei and Pan, Jian-Song and Yan, Bo},
  journal={Light: Science \& Applications},
  volume={14},
  number={1},
  pages={326},
  year={2025},
  url={https://doi.org/10.1038/s41377-025-02025-3},
  doi={10.1038/s41377-025-02025-3}
}

@article{guo2025many,
  title={Many-body interference in kagome crystals},
  author={Guo, Chunyu and Wang, Kaize and Zhang, Ling and Putzke, Carsten and Chen, Dong and van Delft, Maarten R and Wiedmann, Steffen and Balakirev, Fedor F and McDonald, Ross D and Gutierrez-Amigo, Martin and others},
  journal={Nature},
  pages={1--6},
  year={2025},
  url={https://doi.org/10.1038/s41586-025-09659-8},
  doi={10.1038/s41586-025-09659-8}
}

@Article{negi2025magnetic,
author ="Negi, Pranav and Medhi, Koushik and Pancholi, Abhinav and Roychowdhury, Subhajit",
title  ="Magnetic Kagome materials: bridging fundamental properties and topological quantum applications",
journal  ="Mater. Horiz.",
year  ="2025",
volume  ="12",
issue  ="13",
pages  ="4510-4544",
publisher  ="The Royal Society of Chemistry",
doi  ="10.1039/D5MH00120J",
url  ="http://dx.doi.org/10.1039/D5MH00120J"}

@article{zhou2023observation,
  title = {Observation of flat-band localization and topological edge states induced by effective strong interactions in electrical circuit networks},
  author = {Zhou, Xiaoqi and Zhang, Weixuan and Sun, Houjun and Zhang, Xiangdong},
  journal = {Phys. Rev. B},
  volume = {107},
  issue = {3},
  pages = {035152},
  numpages = {8},
  year = {2023},
  month = {Jan},
  publisher = {American Physical Society},
  doi = {10.1103/PhysRevB.107.035152},
  url = {https://link.aps.org/doi/10.1103/PhysRevB.107.035152}
}

@article{disante2026kagomeM,
  title = {Kagome metals},
  author = {Di Sante, Domenico and Neupert, Titus and Sangiovanni, Giorgio and Thomale, Ronny and Comin, Riccardo and Checkelsky, Joseph G. and Zeljkovic, Ilija and Wilson, Stephen D.},
  journal = {Rev. Mod. Phys.},
  volume = {98},
  issue = {1},
  pages = {015002},
  numpages = {53},
  year = {2026},
  month = {Feb},
  publisher = {American Physical Society},
  doi = {10.1103/1g9n-wm38},
  url = {https://link.aps.org/doi/10.1103/1g9n-wm38}
}

@article{yu2025quantumG,
  title={Quantum geometry in quantum materials},
  author={Yu, Jiabin and Bernevig, B Andrei and Queiroz, Raquel and Rossi, Enrico and Torma, Paivi and Yang, Bohm-Jung},
  journal={npj Quantum Materials},
  volume={10},
  number={1},
  pages={101},
  year={2025},
  publisher={Nature}, 
  url={https://doi.org/10.1038/s41535-025-00801-3}, 
  doi={10.1038/s41535-025-00801-3}
}

\end{document}